\documentclass[aps,prl,showpacs,preprintnumbers,amsmath,amssymb,twocolumn]{revtex4}

\usepackage{graphicx}
\usepackage{dcolumn}
\usepackage{bbm}

\newcommand{\be}{\begin{equation}}
\newcommand{\ee}{\end{equation}}
\newcommand{\ba}{\begin{array}}
\newcommand{\ea}{\end{array}}
\newcommand{\bqa}{\begin{eqnarray}}
\newcommand{\eqa}{\end{eqnarray}}

\newcommand{\tr}{\mbox{Tr}}                                        

\newcommand{\ket}[1]{\ensuremath{| #1 \rangle}}
\newcommand{\prj}[1]{\ensuremath{| #1 \rangle \langle #1 |}}
\newcommand{\ovl}[2]{\ensuremath{\langle #1 | #2 \rangle}}

\begin{document}

\title{Decoherence and multipartite entanglement}

\author{Andr\'e R. R. Carvalho, Florian Mintert and Andreas Buchleitner}
\affiliation{Max-Planck-Institut f\"ur Physik komplexer Systeme,
N\"othnitzer Strasse 38, D-01187 Dresden}

\date{\today}

\begin{abstract}
We study the dynamics of multipartite 
entanglement under the influence of decoherence.
A suitable generalization of concurrence
reveals distinct scaling of the entanglement decay rate of GHZ and W states, 
for various environments.
\end{abstract}


\maketitle

The notion of entangled states is, since the early days of quantum 
mechanics, a key concept when it comes to distinguish 
between the quantum and
the classical world. Besides this fundamental aspect, entanglement attracts
considerable interest since it can be viewed as an indispensable ingredient 
for quantum information processing. Numerous experiments
have recently been carried out in this area, in particular on the controlled 
generation
of entanglement between many quantum
systems~\cite{bouwmeester,rauschenbeutel,sackett, pan, roos} -- such
as to accomplish fundamental scalability requirements for quantum computation. 

An obstacle for the production and observation of multipartite
entanglement resides in its fragility under the unavoidable
interaction with the environment.
This challenges the experimentalist, but also demands a proper
theoretical description of multipartite entanglement in open systems.
Despite recent progress in the understanding of decoherence processes
in quantum systems, a systematic characterization of the environment
induced loss of many-particle quantum correlations is still lacking,
and this is specifically due to the difficulties in quantifying
multipartite entanglement.

Even for bipartite mixed states, apart from the particular case of two level
 systems where an exact solution is known~\cite{wootters}, the situation is far from being
 simple. Some of the widely used indicators of entanglement, such as the
 positive partial transpose and negativity~\cite{vidal_pra02}, fail to detect certain entangled
 states, while some other entanglement measures for mixed states require a
 high dimensional optimization procedure which only provides an upper bound,
 unable to reliably distinguish entangled from separable states. Only
 recently~\cite{mintert}, a lower bound (together with a numerically manageable
 upper bound) for the concurrence of mixed bipartite quantum states was
 derived. In the multipartite case, one usually has to deal with bipartite
 cuts, 
where the $N$ individual constituents are partitioned into two (arbitrarily
 chosen) subgroups. The available entanglement measures for bipartite systems
 thus become applicable, though 
different cuts or partitions may lead to different values of 
entanglement and, furthermore, the number of possible bipartite cuts increases
rapidly with $N$. 

To improve on this situation, we will here scrutinize 
the effects of decoherence on multipartite entanglement
measured by a suitable generalization of concurrence -- sensitive to
multipartite correlations. 
Moreover, since different experimental strategies to produce multipartite
entanglement are subject to different sources of decoherence, we implement our
approach for different types of environment coupling, acting on different
types of initially maximally entangled multipartite quantum states. This
finally 
provides a versatile toolbox to assess the scaling of
entanglement decay with the system size under very general conditions.

To start with, we assume that each individual particle of the system interacts
independently with the environment and, therefore, undergoes a local
decoherence process which is mediated by dissipation, noise or dephasing. 
Dissipative effects are described by the coupling of the
system to a thermal bath at zero temperature, and can represent, e.g., 
the spontaneous decay of a two level atom induced by its interaction with
the vacuum modes of the 
ambient electromagnetic field. Noisy dynamics are related to another
limit of the thermal reservoir, when temperature tends
to infinity whilst the coupling strength goes to
zero. Dephasing corresponds to a situation where no energy is exchanged with
the environment, but only phase information is lost. 
All these processes can be described in terms of the master equation   
\begin{equation}
\label{eq:lindblad}
\frac{d{\rho}}{dt}=\sum_{k=1}^N\Bigl(
\underbrace{{\mathbbm 1}\otimes\hdots\otimes{\mathbbm 1}}_{k-1}\otimes{\cal L}_k\otimes
\underbrace{{\mathbbm 1}\otimes\hdots\otimes{\mathbbm 1}}_{N-k}\Bigr)\rho,
\end{equation}
where $ \rho$ is the reduced density operator of the system. The Lindblad
operators ${\cal L}_k$, describing the independent interaction of
each particle $k$ with 
the reservoir, are assumed to be of the same form for all particles and can be
written, in the weak coupling regime and in the Markovian limit, as
\begin{equation}
\label{lindop}
{\cal L}_k  \rho=
\sum_i\frac{\Gamma_i}{2}\left(2\,{c}_i\,{\rho}\,{c}_i^\dagger - 
{c}_i^\dagger\,{c}_i\,{\rho} -
{\rho}\,{c}_i^\dagger\,{c}_i\right)\, ,
\end{equation}
where ${c}_i$ and $\Gamma_i$ describe, respectively, the
 system-environment coupling operator and its strength. 
Further specializing to the case that
 each particle is a two level system (the generic scenario in ion trap quantum
 information processing), the operators $ c_i$
 in~(\ref{lindop}) can be written in terms of the Pauli matrices and are given
 by 
$ c =  \sigma_{-}$, for the zero temperature
 reservoir, by $ c_1=  \sigma_{-}$ and $ c_2=   \sigma_{+}$,
 for the infinite temperature environment, and by $ c 
 =  \sigma_+ \sigma_-$ for dephasing, with  $\Gamma_i = \Gamma$
 in all cases. 

It is clear that, under the action of any of the above environments, any
initially entangled state will asymptotically evolve into a separable state. We are interested in the rate at which multipartite
entanglement decays, and therefore need a suitable measure of
multipartite entanglement which is evaluable for general states. For this
purpose, we use a generalization of the concurrence of pure bipartite states
\cite{run01}, 
$C_2(\Psi)=\sqrt{2(\ovl{\Psi}{\Psi}^2-\tr\varrho_r^2)}$,
where the reduced density matrix $\varrho_r=\tr_p\prj\Psi$ is obtained
as a partial trace of the bipartite state.
For $N$-partite systems, one can construct $2^N-2$ different
reduced density matrices, half of which in general have a different mixing.
A generalization of $C_2$ is therefore not unique, providing several
inequivalent alternatives. For our present purpose, we will focus on the
specific form 
\be
C_N(\Psi)=2^{1-\frac{N}{2}}
\sqrt{(2^N-2)\ovl{\Psi}{\Psi}^2- \sum_\alpha\tr\varrho_\alpha^2}\ ,
\label{multi}
\ee
where $\alpha$ labels 
all different reduced density matrices, {\it i.e.}, there are $N!/(N-n)!n!$
  different terms when $\varrho_\alpha$ is obtained by tracing over $n$ different
  subsystems. $C_N$ vanishes exactly if $\ket{\Psi_N}$ is $N$-separable, {\it
  i.e.} $\ket{\Psi_N}=\bigotimes_{i=1}^N\ket{\Phi_i}$, where the
  $\ket{\Phi_i}$ are pure states of the individual subsystems, 
and $C_N$ adopts its maximal value for GHZ-states
$\sum_i\ket{i\hdots i}/\sqrt{2}$.
Hence, the crucial merit of our specific 
definition
(\ref{multi}) is that $C_N$ can account for real multi-partite correlations \cite{mpc} -- as
opposed to other multipartite generalizations of concurrence
\cite{meyer_jmp02,brennen_jqi03} which extract only bipartite correlations between
single subsystems and the remainder. Also note that eq.~(\ref{multi})
satisfies another 
important requirement for our present purpose -- it allows to compare the degree
of entanglement of multipartite systems with different numbers of constituents:
Consider an $N$-partite state $\ket{\Psi_N}$ that factorizes into an
$N-1$-partite state $\ket{\Psi_{N-1}}$ and a one component 
state
$\ket{\Phi}$, $\ket{\Psi_N}=\ket{\Psi_{N-1}}\otimes\ket{\Phi}$.
In this case, the $N$-partite concurrence $C_N(\Psi_N)$ simply
reduces to the $N-1$-partite concurrence of $C_{N-1}(\Psi_{N-1})$,
and thus allows for a 
meaningful comparison of $N-1$ with $N$-particle states.

To monitor the time evolution of entanglement, we still need 
the generalization of eq.~(\ref{multi}) for mixed states, given as
\be
C_N( \rho)=\inf\sum_ip_iC_N(\Psi_i)\ ,
\label{roof}
\ee
where the infimum is to be found among all sets of probabilities $p_i$
and pure states $\ket{\Psi_i}$, such that
$ \rho=\sum_ip_i\prj{\Psi_i}$.
In principle, this defines an optimization problem that is very
difficult to solve exactly.
Though, lower bounds for $C_N( \rho)$ are available
\cite{mintert},
and can be determined purely algebraically
in the regime where the mixing of $\rho$ is moderate \cite{mintert_phd,flo_qp}.
Moreover, for states of rank two -- that we 
shall encounter for specific time evolutions considered further down -- 
eq.~(\ref{roof}) can be condensed into an expression 
equivalent to the one for
bipartite two-level systems \cite{wootters,mintert},
and the required optimization 
can be performed exactly.

With the definition (\ref{roof}) at hand, we can monitor the multipartite
concurrence $C_N$ 
for the solutions $ \rho(t)$ of eq.~(\ref{eq:lindblad}). We will consider
two types of initial states: 
the GHZ state, $\vert
\Psi_{N}\rangle_{\rm GHZ}=(\vert 00 \hdots 0 \rangle + \vert 11 \hdots 1
\rangle)/\sqrt{2}$, 
and the W state, $\vert
\Psi_{N}\rangle_{W}=(\vert 00 \hdots 01 \rangle + \vert 00 \hdots 10
\rangle + \hdots +\vert 10 \hdots 00 \rangle)/\sqrt{N}$ -- which are known to
bear incompatible multipartite correlations, in the sense that they cannot be
transformed into each other by local unitary transformations amended by
classical communication \cite{duer00}. 
Also note that these states have
recently been produced \cite{sackett,roos} for $N=3,4$ in the lab,
and that it is now within experimental reach 
to monitor the time dependence of their degree of
entanglement by means of quantum state tomography. 

\begin{figure}
\includegraphics[width=8.0cm]{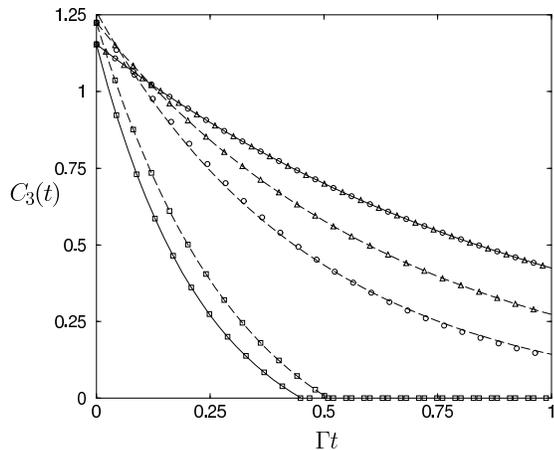}
\caption{Time dependence of the multipartite concurrence $C_N(t)$, for a
$N=3$-partite system initially prepared in
a GHZ (dashed lines) or W (solid lines) state. Only for an infinite
temperature reservoir (squares) does the state turn separable after a finite
time $t_{\rm sep}$. In contrast, the  zero temperature (circles) as well as
the dephasing environment (triangles) induce separability only in the limit
$t\rightarrow\infty$. In all cases, the numerical results are very well fitted
by an exponential (solid and dashed lines).}
\label{decay}
\end{figure}

As a first observation, illustrated in Fig.~\ref{decay}, we find 
an essentially perfect monoexponential decay of concurrence 
in all analyzed cases, with a decay rate $\gamma$ which depends on the initial
condition and environment model. Only for an infinite temperature reservoir
does $C_N(t)$ vanish after a finite time $t_{\rm sep}$, as illustrated in
Fig.~\ref{decay}, consistent with the results obtained in \cite{kempe,briegel}
for a depolarizing channel. In contrast, for zero temperature as well as for
dephasing environments does $C_N(t)$ only vanish for $t\rightarrow\infty$.
In these two cases the concurrence of the decaying W-state gets larger than
that of the decaying GHZ-state after a short time.
In contrast to this, the situation of the initially prepared states, where the
concurrence of the GHZ-state is larger than that of the W-state, is preserved
for the infinite temperature reservoir.
To allow for a consistent comparison of the decay
of entanglement in arbitrary environments, and, furthermore, since experimentalists seek 
to maximize and to preserve entanglement by preparing
pure entangled states and minimizing decoherence, rather than to await its
fading out, we therefore choose $\gamma$ rather than $t_{\rm
  sep}$ as the figure of merit in our analysis of entanglement
decay for variable system size.

Consistently with the experimental scenario, we
assume that the initial multipartite states are pure (or quasi pure, i.e. the
experimental preparation of the initial quantum state succeeds with
essentially perfect fidelity) \cite{sackett,roos}.
Thus, on short time scales, where the degree of mixing of $ \rho$ is
expected to be small, the use of our quasi-pure approximation~\cite{flo_qp} 
is justified. Moreover, we explicitely verified that, even for considerably
mixed states, this approximation provides an excellent estimate of $C_N$ in eq.~(\ref{roof}) with significantly reduced computational effort. 
Since we always evaluate lower bounds on concurrence, all decay rates shown
hereafter present an {\em upper} bound of the actual rate at which
multipartite entanglement is lost due to contact with the environment.

\begin{figure}
\includegraphics[width=7.5cm]{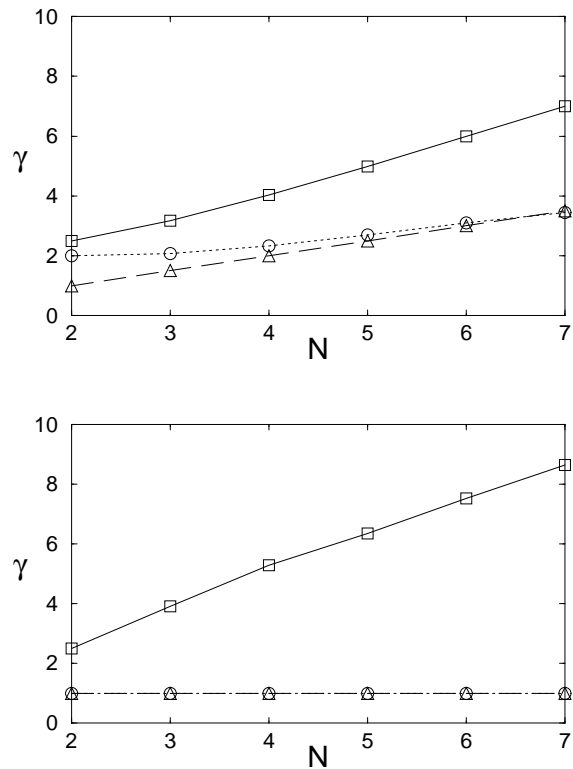}
\caption{Decay rates $\gamma$ (in units of the reservoir rate $\Gamma$) for
GHZ (top) and W (bottom) states, as a function of the system size (i.e., 
the particle number)
$N$. The different 
environment models are represented by circles connected by a dotted line (zero
temperature), by squares connected by a solid line (infinite temperature), and
by triangles connected by a 
long-dashed line (dephasing), respectively. Whilst for GHZ states the
decay rates increase roughly linearly with $N$, independently of the specific
environment,
the W states exhibit increasing decay with system size 
only for the infinite temperature environment. Remarkably, the decay rate of
the W states is size-independent for dephasing and zero temperature environments!}
\label{gammas}
\end{figure}

Because of their primordial experimental interest, we
focus on the multipartite entanglement of GHZ and W states of variable size. 
Figure~\ref{gammas} shows the scaling of their entanglement decay rates 
$\gamma$ with $N$, 
under decoherence induced by zero temperature, infinite
temperature, and dephasing environments. We see that the GHZ state (top panel
of the figure) decays into
a separable state with a rate which increases linearly with $N$, except for
the small-$N$ behavior 
of $\gamma$ for the zero temperature
environment.
Indeed, the special case of a dephasing reservoir, 
where the density matrix is always a mixture of two pure states, and, hence,
is of rank two,
can be treated analytically. 
In this particular case, 
the concurrence just follows the behavior of
the two non-vanishing 
non-diagonal elements of $ \rho$ which uniformly decay as $e^{-N\Gamma
  t/2}$.

Remarkably, 
the situation changes quite drastically for the W states (bottom plot of
Fig.~\ref{gammas}). In this case, only the infinite temperature environment 
gives rise to an almost linear increase of $\gamma$ with $N$, slightly faster
than for the GHZ states. In contrast, for dephasing and zero temperature
reservoirs, the decay of the concurrence is {\em independent} of
$N$. Moreover, the zero temperature case once again allows for an analytic solution (as above, the rank of the state reduces to two) for all $N$, 
leading to $C_N(t)\sim e^{-\Gamma t}$.
Consequently, the multipartite quantum
correlations of W states clearly outperform those of GHZ states in terms of
their robustness.
One might be tempted to attribute this to the smaller initial concurrence
of W as compared to GHZ states.
Though, the ratio
\begin{equation}
\frac{C_N(\Psi_{GHZ})}{C_N(\Psi_W)}=
\sqrt{(1-2^{1-N})\frac{N}{N-1}}\nonumber
\end{equation}
with maximum for $N=5$, approaches unity for large $N$.

To summarize, we have shown that an efficient monitoring of multipartite
entanglement under arbitrary (Markovian) environment coupling is possible, for
different classes of quasi pure initial states typically dealt with in state-of-the-art experiments. Furthermore, our finding that GHZ states are
significantly more
fragile under environment coupling than W states might indicate a robust
pathway to scalable quantum information processing.

\bibliography{multpartdeco}

\end{document}